\documentclass[preprint2]{aastex}

\usepackage{graphicx}
\usepackage{psfig}
\usepackage{epsfig}
\usepackage{amsmath}

\newcommand{\lsun}{\,\hbox{$L_{\odot}$}}

\newcommand{\reff}{\mbox{$R_{\rm eff}$}}

\newcommand{\lir}{\,\hbox{$L_{\rm IR}$}}
\newcommand{\lopt}{\,\hbox{$L_{\rm opt}$}}
\newcommand{\hst}{{\it Hubble Space Telescope}}
\newcommand{\spi}{{\it Spitzer}}
\newcommand{\ang}{\,\hbox{\AA}}

\shorttitle{NICMOS imaging of z$\sim$2 ULIRGs}
\shortauthors{Dasyra et al.}

\begin{document}

\title{{\it \bf{HST}} NICMOS Imaging of z$\sim$2, 24~$\micron$-selected Ultraluminous Infrared Galaxies}

\author{Kalliopi M. Dasyra\altaffilmark{1}, Lin Yan\altaffilmark{1}, 
George Helou\altaffilmark{1}, Jason Surace\altaffilmark{1}, 
Anna Sajina\altaffilmark{1}, James Colbert\altaffilmark{1}}

\altaffiltext{1}{Spitzer Science Center, California Institute of Technology,
Mail Code 220-6, 1200 East California Blvd, Pasadena, CA 91125}

\begin{abstract}
We present \hst\ NICMOS $H$-band imaging of 33 ultraluminous infrared galaxies (ULIRGs) 
at $z\sim$2 that were selected from the 24 \micron\ catalog of the \spi\ Extragalactic 
First Look Survey. The images reveal that at least 17 of the 33 objects are associated 
with interactions. Up to one-fifth of the sources in our sample could be minor mergers, 
whereas only two systems are merging binaries with luminosity ratio $\le$3:1, which is 
characteristic of local ULIRGs. The rest-frame optical luminosities of the sources are of 
the order $10^{10}$-$10^{11}$ \lsun\ and their effective radii range from 1.4 to 4.9 kpc. 
The most compact sources are either those with a strong active nucleus continuum
or those with a heavy obscuration in the mid-infrared regime, as determined from \spi\ 
Infrared Spectrograph data. The luminosity of the 7.7\micron\ feature produced by 
polycyclic aromatic hydrocarbon molecules varies significantly among compact systems,
whereas it is typically large for extended systems. A bulge-to-disk decomposition 
performed for the six brightest ($m_H$$<$20) sources in our sample 
indicates that they are best fit by disk-like profiles with small or negligible bulges, 
unlike the bulge-dominated remnants of local ULIRGs. Our results provide evidence that 
the interactions associated with ultraluminous infrared activity at $z\sim$2 can differ 
from those at $z\sim$0.

\end{abstract}

\keywords{
infrared: galaxies ---
galaxies: evolution ---
galaxies: formation ---
galaxies: interactions ---
galaxies: high-redshift
}


\section{Introduction}
\label{sec:intro}

The excess of infrared (IR) emission in extragalactic sources is widely believed 
to be triggered most efficiently by gas-rich galactic mergers or interactions
(e.g, \citealt{larson}; \citealt{sanders88a}). Dynamical instabilities are required 
to drive large concentrations of gas at small radii and to lead to a high rate of 
star-formation and an excess of IR emission ($>10^{11}$ \lsun ). The number 
density of both IR-bright sources and interactions evolves strongly with redshift 
$z$, increasing to a redshift of at least 1 (\citealt{elbaz02}; 
\citealt{lefloch05}; \citealt{perez}; \citealt{conselice}; \citealt{jeyhan}). 
However, it is still unclear whether the interaction mechanisms that lead to 
high IR outputs are the same in the local Universe and at high redshifts. 

Observations of ultraluminous IR galaxies (ULIRGs; $L_{8-1000\mu m}>10^{12}L_\odot$) 
in the local Universe indicate that ULIRGs represent a well-defined merger category.
They are triggered by the merger of two gas-rich galaxies whose typical gas fractions 
are $\sim$10\% of their dynamical masses (\citealt{downes}).  The merging components
have comparable masses (\citealt{dasyra06a}) or luminosities (\citealt{kim02}; 
\citealt{murphy96}). The fairly specific origin of ULIRGs is thought to be related to 
their high phase-space gas densities, which are achieved through major mergers (e.g., 
\citealt{barnes}). The merger typically leads to the formation of elliptical galaxies  
(\citealt{joseph}). The remnants often have light profiles that follow an r$^{1/4}$ 
distribution (\citealt{kormendy};  \citealt{veilleux02}), fall on the velocity 
dispersion $-$ effective radius \reff\ projection (\citealt{genzel01}; 
\citealt{tacconi02}; \citealt{dasyra06b}) of the fundamental plane of early-type 
galaxies (\citealt{dressler}; \citealt{djoda}), and have a bulge-to-disk ratio 
of $\sim$1.5 (\citealt{veilleux06}).  

The trigger mechanisms and end products of lower luminosity systems, such as the local 
luminous IR galaxies (LIRGs;  $L_{8-1000\mu m} \sim 10^{11}-10^{12}L_\odot$), are not 
constrained as well as those of local ULIRGs. 
The imaging analysis of \cite{alonso-herrero} indicates that most local LIRGs have 
prominent spiral patterns, and of those, a nonnegligible number are weakly interacting
or even isolated systems. \cite{ishida} found that several minor mergers or fly-by 
interactions exist at the low-luminosity end of local LIRGs. However, as the luminosity 
increases so does the fraction of major mergers. Some LIRGs need to have the same 
origin as ULIRGs since every ULIRG has to pass through a LIRG phase before and after 
its main starburst episode (\citealt{murphy}). 

At high redshifts the situation is not as well defined, mostly because the merging 
galaxies are on average more gas rich than those in the local Universe. The amount of gas 
in, e.g., submillimeter galaxies is $\sim$40\% of their total mass (\citealt{tacconi06}). 
It has been suggested that a variety of galaxy types can be produced by mergers of high 
gas fractions. The simulations of \cite{springelred} and \cite{springelspiral} 
indicate that the merger outcome can vary between a red and dead elliptical 
or a spiral galaxy, depending on how much gas is converted into stars or expelled 
from the galaxy. \cite{hammer} suggest that most local spiral galaxies may have 
undergone mergers within the last 8 Gyr by invoking star-formation duration 
and mass assembly arguments. 

The goal of this paper is to investigate what processes lead to an ultraluminous 
IR activity at redshifts 1.5$<z<$2.5. Could high-$z$ ULIRGs be triggered in an analogous 
manner to local ULIRGs or to local LIRGs? If triggered by interactions, what are their 
progenitors and their end products? 
To address these questions, we requested observations of a sample of 35 ULIRGs at 
$z\sim$2 using the Near Infrared Camera and Multi-Object Spectrometer (NICMOS) 
instrument on board the {\it Hubble Space Telescope} ({\it HST}) .

The rest of this paper is organized as follows. In \S~\ref{sec:sample} 
and \S~\ref{sec:reduction} we present the sample and the data reduction used for the
analysis in \S~\ref{sec:analysis}. The presentation of the results in \S~\ref{sec:host}
is followed by a comparison between z$\sim$2 and local ULIRGs in \S~\ref{sec:comp} and 
the conclusions in \S~\ref{sec:conc}. Throughout this paper we use a $\Lambda$CDM cosmology 
with $H_0$=70~km~s$^{-1}$~Mpc$^{-1}$, $\Omega_{m}$=0.3, and $\Omega_{\Lambda}$=0.7. 


\section{The Sample}
\label{sec:sample}

\begin{figure*}
\centering
\includegraphics[width=7cm]{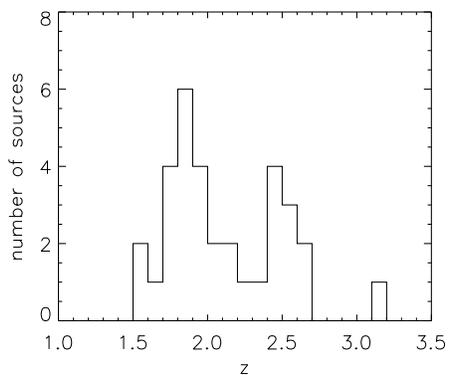}
\caption{\label{fig:z_hist} Redshift distribution of the ULIRG sample
observed with the \hst\ NIC2 camera.}
\end{figure*}

Our sources are drawn from a sample of IR-bright galaxies detected in the 24 
\micron\ mosaic of the 3.7 deg$^2$ \spi\ 
Extragalactic First 
Look Survey (XFLS; \citealt{fadda}). The selected sources have 24 \micron\ 
fluxes greater than 0.9 mJy (\citealt{yan05}). Two extra criteria were imposed 
to ensure that the sample includes high-$z$ starburst galaxies: the flux 
ratios {log($\nu f_{\nu}$[24 \micron ]/ $\nu f_{\nu}$[8 \micron ])} and
{log($\nu f_{\nu}$[24 \micron ]/ $\nu f_{\nu}$[6440\,\ang ])} were required 
to be greater than or equal to 0.5 and 1, respectively. The use of these
threshold values is based on template spectral energy distributions 
(SEDs) of local starbursts (\citealt{yan04}). The high IR-to-optical flux ratio 
preferentially selects starbursts at $z\ge0.6$, since the observed $R$-band 
magnitudes sample the SEDs shortward of the 4000\ang\ break. The high 24/8 \micron\ 
flux ratio can either pick strong starbursts at $1.5<z<3$  whose 6.2, 7.7, and 8.6 
\micron\ polycyclic aromatic hydrocarbon (PAH) features are shifted in the 24 
\micron\ beam or obscured, reddened active galactic nucleus (AGN) continua (Yan 
et al. 2004, 2007; \citealt{brandl}). In total, 52 XFLS sources that satisfy 
these criteria were observed with the Infrared Spectrograph (IRS; \citealt{irs}) 
on board \spi . The mid-infrared (MIR) spectra yielded redshifts for 47 sources 
in the range [0.61,3.2] (\citealt{yan07}). For the NICMOS observations, we 
chose to observe only the sources at $z>1.5$ so that we can perform an imaging 
analysis at comparable rest-frame wavelengths for all sources. Of the 35 
sources at $z>1.5$, 33 were successfully observed. These 33 sources are presented
in Table~\ref{tab:obs} and their redshift distribution is presented in 
Figure~\ref{fig:z_hist}.

From the IR point of view, most of the sources in this sample are ULIRGs. The rest-frame 
14 \micron\ luminosity, L$_{14}$, which is based on spectral fitting of IRS 
data by \cite{sajina07}, is greater than 10$^{12}$\lsun\ for 27 sources, indicating that
their bolometric IR luminosities, \lir , are in the range of 
ULIRGs or higher. The bolometric IR luminosities are computed, when possible,
by fitting the far-infrared (FIR) SEDs of the sources using \spi\ MIPS 70 
and 160 \micron\ fluxes, and Max-Planck Millimeter Bolometer 1.2 mm fluxes 
(Sajina et al. 2008; see Table~\ref{tab:obs}).  
The SED fitting indicates that three more sources have \lir\ values that exceed the 
ULIRG threshold. On average, the IR luminosity of these sources is substantially 
greater than that of local ULIRGs (e.g., \citealt{kim02}) and is close to the 10$^{13}$ 
\lsun\ limit, which defines the hyperluminous IR galaxies. The remaining three 
sources, namely MIPS~8493, MIPS~16133, and MIPS~22661 are also likely to be ULIRGs 
given their high $L_{14}$ values and \lir\ upper limits (Sajina et al. 2008).

The 7-38 \micron\ IRS spectra revealed that the IR emission arises to comparable
levels from intense star formation, AGN activity, or a combination of the 
two (\citealt{yan07}). About one-third of the sample has a strong silicate absorption 
feature at 9.7 \micron, which implies that the main emission mechanism is compact and 
heavily obscured; it can either be a nuclear starburst and/or an AGN (\citealt{spoon}). 
Therefore, the IR emission of high-$z$ ULIRGs has a multiple origin, similar to that of 
local ULIRGs as revealed by {\it Infrared Space Observatory} (ISO; \citealt{genzel98}; 
\citealt{rig99}) and \spi\ spectroscopic data (\citealt{armus}). 


\section{Observations and data reduction}
\label{sec:reduction}
The observations of this program (ID 10858) were carried out using 
the NIC2 camera of the NICMOS instrument. The pixel scale of NIC2 corresponds 
to 0\farcs076 and 0\farcs075 along the x and y axis respectively. The 
resolution of the images is high, 0\farcs158, as measured from the full 
width half maximum (FWHM) of stellar light profiles. The images were taken
in the $H$-band using the F160W filter, which is centered at a wavelength  
of 1.6 \micron. The wavelength that corresponds to the center of the filter 
at the rest frame of each source is in the range [4310, 6150] \ang.  The 
width of the F160W filter at the rest frame of each source is presented in
Table~\ref{tab:obs}. The detector read-out mode was set 
to ``multiaccum'' with 23 or 24 consecutive, non-destructive reads leading 
to an individual exposure. The observations were performed using a spiral 
four-point dither pattern per orbit to optimize the removal of bad pixels. 
The total on-source integration time for the four dither positions was either 
2688 or 2560 s. The number of orbits per source was determined by the source's 
\spi\ Infrared Array Camera (IRAC; \citealt{irac}) 3.6 \micron\ flux. The 
sources with flux density greater than the detection limit of the XFLS 
catalog (20 $\mu$Jy; \citealt{lacy05}) were typically observed for a single 
orbit. The sources that are fainter than this threshold were observed for two orbits. 
The total on-source integration times are presented in Table~\ref{tab:obs}.
 
\begin{figure*}
\centering
\includegraphics[width=16cm]{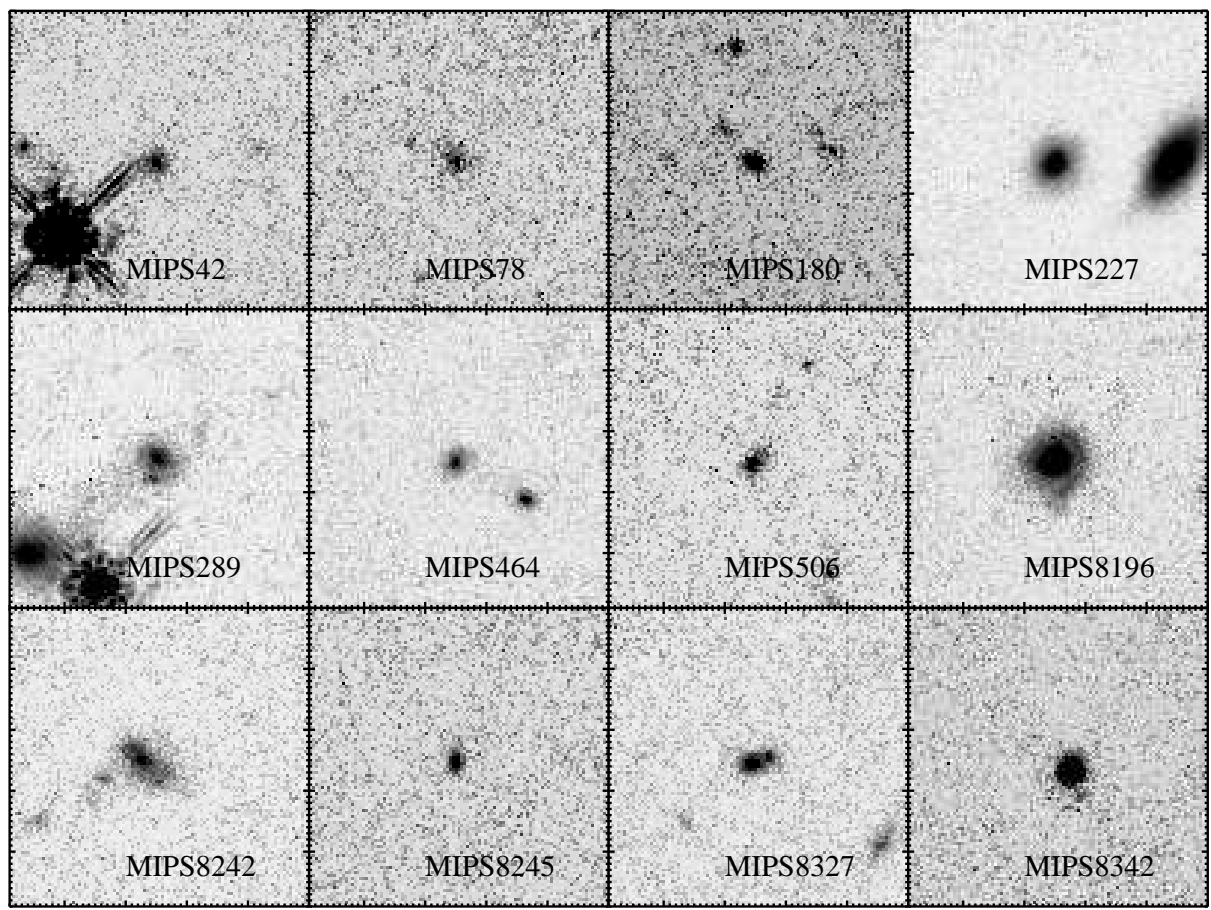}
\caption{\label{fig:images} Reduced NIC2 images of the z$\sim$2 ULIRGs in
our sample. All the images in this figure correspond to a 
$3\farcs8 \times 3\farcs8$ field of view and are displayed in a squared
scale.} 
\end{figure*}

The reduction and calibration of the NICMOS images were performed using
standard {\it HST} IRAF and IDL routines and following the prescriptions of
the NICMOS manual (\citealt{nicmos}). The pipeline {\it calnica} function 
was used to remove the zeroth detector read-out image from the non-destructive 
consecutive read-out images, and to correct for detector nonlinearities, 
dark currents, and bar effects. The same task was also used to flat 
field and combine the consecutive read-out images into a single image, computing 
its statistical error. We then used the IRAF task {\it pedsky} to bring the 
quadrants of the detector to similar bias levels and to apply a 
zeroth order removal of the sky background. Following {\it pedsky}, we 
corrected for South Atlantic anomaly effects using the IDL procedure 
{\it saa clean}. To correct for systematic sky residuals and enhance the signal-to-noise 
ratio of our background-limited observations, we constructed a residual 
sky image by median combining the images at the different dither positions. 
We then removed the sky residuals from each of these images, shifted them into 
their common reference position, and removed cosmic rays and bad pixels. 
The fluxes of the bad pixels were replaced with their average value in 
all other useful dither positions. To enhance the resolution of our images, we 
resampled the pixels to half their original size using the IRAF routine 
{\it drizzle}. This routine simultaneously corrects for detector geometric 
distortions. The resulting images were weight-averaged using their uncertainty 
images, which were also corrected for shifting and drizzling transformations.
The weight-averaged image and its final uncertainty image were drizzled back 
to their initial pixel scale. The final images are presented in Figure~\ref{fig:images}.

\section{Data analysis}
\label{sec:analysis}

Our images show that a significant fraction of  
sources have irregular patterns indicative of interactions. For the 
proper photometric calibration of such sources, we need to ensure that all 
of their light will be accounted for, even if it is distributed in 
asymmetric structures (e.g., tidal tails). For this reason, we performed a 
basic photometric analysis by fitting a series of isophotal ellipses to 
each galaxy with the {\it sextractor} software (\citealt{sextractor}). 
Using {\it sextractor}, we first ensured the detection of each source at 
a 3$\sigma$ level and then integrated its flux down to 1$\sigma$ level.
{\it Sextractor} measured the flux, the effective (half-light) radius, the 
ellipticity, and the position angle of each galaxy within that detection 
limit. We converted the integrated galaxy counts per second to an apparent 
Vega magnitude using the NICMOS data calibration prescription 
(\citealt{dickinson})
\begin{equation}
\hbox{$m=-2.5 $log($f_{\rm CPS} \times {\rm PHOTFNU} / f_\nu$ [Vega])},
\end{equation}
where f$_{\rm CPS}$ is the source flux in counts per second, PHOTFNU is the 
conversion factor from counts per second to janskys, and $f_{\nu}$(Vega) is 
the Vega flux in janskys for the given NICMOS camera and filter passband. 
PHOTFNU is computed by {\it calnica} and equals 1.49585$\times 10^{-6}$ 
Jy s count$^{-1}$. For NIC2 at F160W, $f_\nu$(Vega) is 1043.6 Jy.

Following the global photometric characterization of all the sources in our 
sample, we decomposed the brightest systems into a bulge and a disk component 
using {\it galfit}\footnote{{\it Galfit} is thought of as the optimal 
code for the extraction of morphological parameters of unrelaxed
galaxies at high redshifts (\citealt{haussler}).} (\citealt{galfit}).
{\it Galfit} implements a $\chi^2$ minimization to decompose the light profile 
of a source into a (set of) predefined function(s). We selected the 
S\'ersic profile, whose intensity
scales with radius as exp$([r/\reff]^{1/n})$. We used two S\'ersic components, 
one with $n$=4 and another with $n$=1, corresponding to a de Vaucouleurs and an 
exponential disk profile respectively. {\it Galfit} convolves both components 
with the instrumental point-spread function (PSF) before simultaneously fitting 
them to the observations. We also performed a PSF versus free S\'ersic index 
decomposition as a consistency check. We simulated the required NIC2 PSF with the 
TinyTim code (\citealt{tinytim}) using a pixel scale identical to that of the 
observations. To perform the $\chi ^2$ minimization, {\it galfit} requires an 
error image. For this purpose, we provided the final uncertainty image.  
{\it Galfit} also requires initial guesses for the 
parameters under examination; we used the output parameters of {\it sextractor }
as input values for the magnitude, the half-light radius, the position angle, 
and the ellipticity of each source. Finally, we masked out all other bright sources 
within each image so that {\it galfit} can properly compute its background level.


\section{Results} 
\label{sec:host}

\subsection{Rest-frame Optical Galaxy Morphologies And Luminosities} 

\begin{figure*}
\centering
\includegraphics[width=12cm]{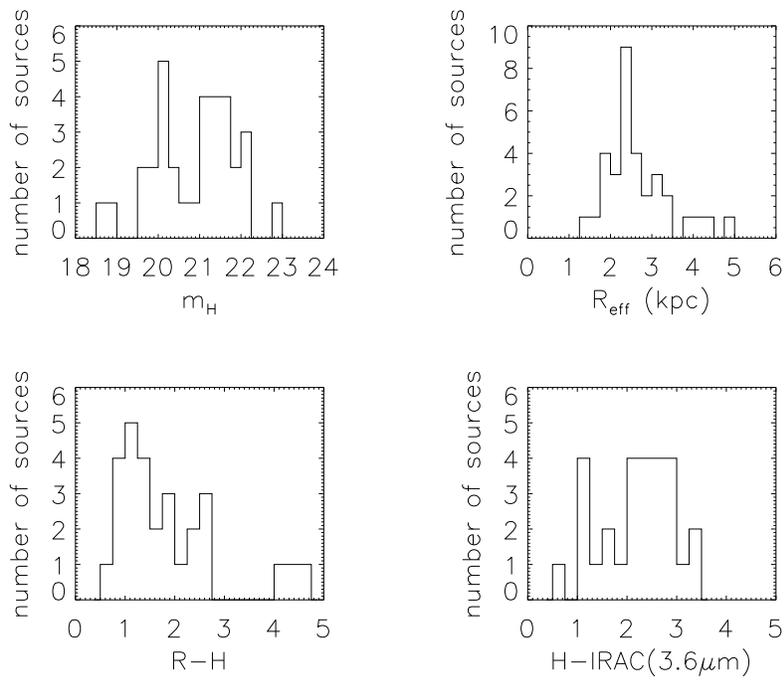}
\caption{\label{fig:m_r_hist} Histograms of the apparent magnitude distribution 
({\it top left}), the effective radius distribution ({\it top right}), the $R-H$ 
colors ({\it bottom left}), and the $H-$IRAC(3.6 \micron ) colors ({\it bottom right}) of 
the sources in our sample.}
\end{figure*}

The mean and median $H$-band (or rest-frame optical) magnitudes of the sources in 
our sample are $m_H$=20.89 and $m_H$=21.07 with a dispersion of 0.98 in Vega 
magnitudes (Table~\ref{tab:phot}; Fig.~\ref{fig:m_r_hist}). For comparison, one of 
the brightest ULIRGs in the local Universe, Mrk 1014, would have $m_H$=20.7 at $z$=2 
(\citealt{kim02}). The distribution of the half-light radii is also presented in 
Figure~\ref{fig:m_r_hist}. The mean and median values of the \reff\ distribution equal 
2.66 and 2.43 kpc, respectively, with a standard deviation of 0.80 kpc. The $R-H$ and 
$H-$IRAC(3.6 \micron ) color distributions (Fig.~\ref{fig:m_r_hist}) have mean values 
of 1.86 and 2.20 and median values of 1.64 and 2.38, respectively. Their standard 
deviations equal 1.06 and 0.74. The surface-brightness $\mu_H$ distribution has a mean, 
median, and standard deviation of 20.33, 20.62, and 1.06 mag arcsec$^{-2}$ 
(Table~\ref{tab:phot}). We computed the optical luminosity \lopt\ of each source at 
the rest-frame wavelength that corresponds to the center of the F160W filter (1.6 \micron ) 
by multiplying its flux with the frequency at the center of the filter. We find that 
the rest-frame optical luminosities range between 1.70 $\times{10^{10}}$ \lsun\ and 
9.46 $\times{10^{11}}$ \lsun\ (Table~\ref{tab:phot}). The average and median luminosity 
values are 1.12 $\times{10^{11}}$ \lsun\ and 7.83 $\times{10^{10}}$ \lsun\, with a wide 
dispersion that is equal to 0.37 dex when measured on a logarithmic scale. 
The rest-frame $L_{14}/\lopt\ $ ratios are in the range [4,218]. For the six sources 
with determined IR luminosities, the $\lir / \lopt\ $ ratios range between 
50 and 281. 

The NIC2 images show that 2 of the 33 sources (MIPS~16059 and MIPS~22530) 
have two merging components of luminosity ratio $\le$3:1. There are three more 
systems, MIPS~289, MIPS~8327, and MIPS~15928, with a secondary faint component 
inside the envelope of their primary nucleus. The faint component could either 
be another nucleus or a star-forming cluster. Other signs of interactions, 
such as perturbed morphologies, stellar fans (MIPS~22558), and tidal tails,
exist in 12 more sources. MIPS~8196 has a bright tail of 19.3$\pm3.7$ kpc.
In total, 52\% of the sources in our sample appear to be involved in interactions. 

This fraction is likely to be a lower limit, given observational limitations. 
Wide binaries that lack prominent signs of interaction or close binaries that 
are unresolved may exist in our sample. In the case of unresolved systems, their 
maximum projected nuclear separation $d_n$ is determined by the instrumental angular 
resolution, which equals 2.1 pixels or 1.34 and 1.28 kpc at redshifts of 1.5 and 2.5,
respectively. Such a resolution is comparable to that of ground-based observations that
were used to study large numbers of ULIRGs in the local Universe (e.g., \citealt{kim02};
\citealt{veilleux02}), which was $\sim$1.8 kpc at z=0.1. Higher angular resolution imaging studies
have shown that a small number of local ULIRGs are likely to have two nuclei separated
by less than $\sim$1.5 kpc (Scoville et al. 1998; 2000; \citealt{bushouse}; \citealt{dasyra06b}).
Flux or surface-brightness detection limitations could also contribute to this effect.
For a single orbit, the flux 3$\sigma$ detection limit of our images corresponds to 
$m_H$=26.07 for a point source and to $m_H$=24.51 for an extended source with a radius 
equal to the average radius of the sources in our sample. This value implies that for a 
primary nucleus of $m_H$=22.00, the secondary nucleus needs to have a 
$\Delta m\lesssim$2.5 in order to be detected. Such a magnitude difference corresponds 
to a luminosity ratio of 10:1. Therefore, small companions in mergers of high luminosity 
ratio could be undetected. Moreover, low surface brightness 
structures that have been diluted into the sky background could exist. The 
surface brightness 3$\sigma$ detection limit of our observations is 22.59 
mag arcsec$^{-2}$ for a single orbit. Tidal tails in local ULIRGs have $B$-band 
surface brightness values between 21.5 mag arcsec$^{-2}$, e.g., in IRAS 22491-1808 
and IRAS 14348-1447, and 23.5 mag arcsec$^{-2}$, e.g., in UGC 5101 (\citealt{surace00a}; 
\citealt{surace00b}).

To further investigate the effects of surface brightness dimming on our results,
we created a model that simulates $z$=2 galaxies from local templates.
The chosen templates, the methodology, and a detailed description of the results 
can be found in the Appendix. We find that the initial light profile is usually 
maintained for sources with $m_H$=19. Extended structures are typically diluted 
into the continuum for sources with $m_H$$>$20. It is therefore likely that a 
nonnegligible fraction of our sources could have diluted low surface brightness features. 
Up to half of the flux can be diluted into noise for sources with $m_H$=21, leading to 
effective radii that can be underestimated by up to a factor of 2. Since the average and 
median effective radius of the local ULIRGs in the 1 Jy sample are 4.85 and 4.70 kpc, 
respectively (\citealt{veilleux02}), the intrinsic radial extents of these 
z$\sim$2 ULIRGs are likely to be comparable to those of local ULIRGs.

In all computations we have ignored any possible extinction or $k$-corrections 
in the rest-frame wavelength range [4200,6200]\ang . \cite{taylor}, who performed 
imaging of a sample of 142 late-type and irregular or interacting galaxies, found that 
their average $B-V$ color is 0.5, corresponding to a 37\% difference in the flux of 
the two bands. This difference in flux does not necessarily correspond to a systematic 
difference in \reff\ from the $B$ to the $V$ band. Often, large color gradients between the 
center and the outer regions of ULIRGs exist (\citealt{surace98}) because the geometry 
of the obscuring medium is complex or highly nucleated (\citealt{soifer02}) due 
to dynamical instabilities. Moreover, accounting and correcting for $B-V$ color differences
is not straightforward since the correction can significantly deviate from its average
value in individual objects. Spectroscopy of local ULIRGs has shown that the continuum 
slope in the optical wavelengths varies from source to source because of different 
AGN-emission strength (\citealt{armus89}; Veilleux et al. 1995; 1999). The broad-band 
colors of nearby filters may also significantly differ due to bright emission lines 
within the filter bandpasses. 

\subsection{Bulge-to-Disk Decomposition Of Bright Sources} 

We used {\it galfit} to find the stellar distribution in $z \sim$ 2 ULIRGs.
We performed this analysis only for the six systems with $m_H$$<$20, because the
light profiles of sources with $m_H \ge $20 can be diluted in the background, leading
to unreliable decomposition results. To infer this reliability limit, we ran 
{\it galfit} for two local disk- and bulge- dominated systems and their simulated z=2 
images. This analysis is presented in detail in the Appendix.  

\begin{figure*}
\centering
\includegraphics[width=16cm]{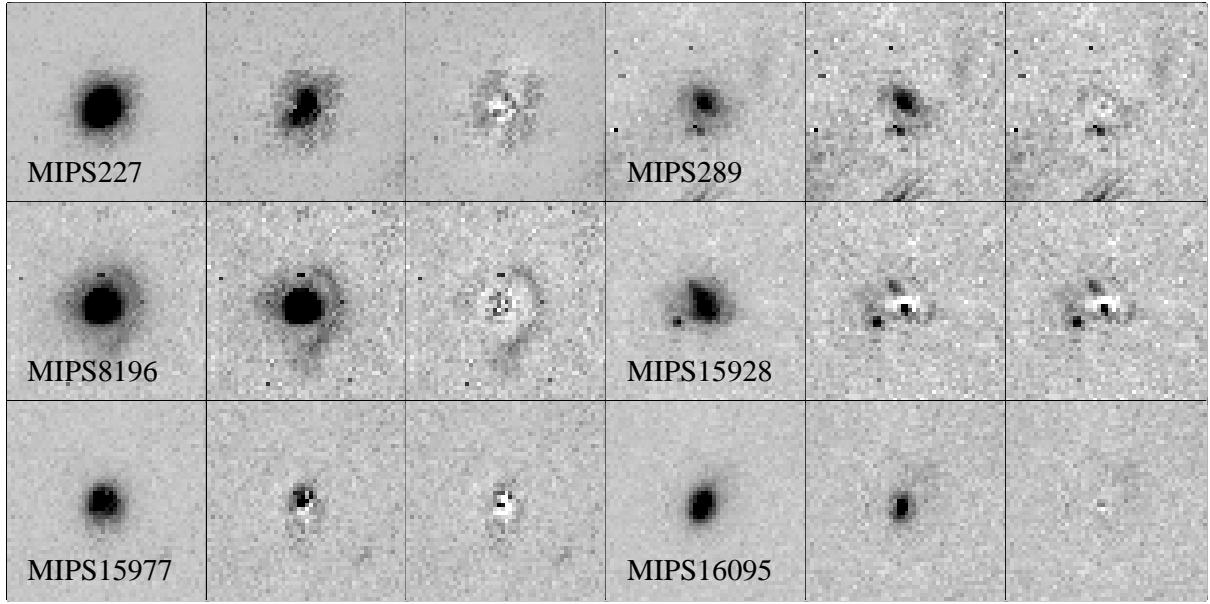}
\caption{\label{fig:residuals} Bulge-to-disk decomposition results for the
bright, extended sources in our sample. For each source, the first panel
corresponds to the original image viewed in a $1\farcs 9 \times 1\farcs 9$
frame. The second panel corresponds to the residuals after the subtraction
of the disk component. The final, bulge-subtracted residuals are presented
in the third column. Whereas the decomposition steps are presented here 
individually for the sake of clarity, the actual {\it galfit} best-fit solution 
was found by simultaneously solving for the optimal bulge and disk parameters. 
} 
\end{figure*}

The bulge-to-disk decomposition results are presented in Figure~\ref{fig:residuals}
and in Table~\ref{tab:decomposition}. The residual flux of the decomposition, which 
on average corresponds to 4.44\% of the initial galaxy flux, is also presented in 
Table~\ref{tab:decomposition}. The decomposition reveals that if 
bulges indeed exist in the sources that we analyzed, they are small. Among the six 
sources, four have a possibly unresolved bulge. This means either that the bulge is 
real but its size is smaller than our resolution element or that it does not 
exist and {\it galfit} identifies a compact nuclear starburst or an AGN as a bulge. 
The mean bulge-to-disk ratio of these sources is 0.62. The median is 0.55, and 
it is significantly smaller than that of local single-nucleus ULIRGs, which is 
1.51 (\citealt{veilleux06}). The measured bulge-to-disk ratios may be 
upper limits of their intrinsic values, since the surface brightness dimming typically 
affects more the disk- than the bulge- dominated light profile distributions  (see 
the Appendix). These results imply that most of these systems had not formed big bulges 
as of $\sim$10 Gyr ago. The exception is MIPS~8196, which is the brightest
system in our sample with $\nu L_{\nu}=9.46\times 10^{11}$ \lsun\ and a 
bulge-to-disk ratio $>$1.

Since a small bulge can be degenerate with a PSF (\citealt{simmons08}) in 
decomposition algorithms, we also performed 
a PSF versus free S\'ersic index $n$ decomposition to check the consistency of our
results. We find that the effective radii of the free-index component agree
on average to 8.53\% with those computed using SExtractor. We also find that 
the PSF flux is always fainter than the galaxy flux by more than a factor of 3 
(or 1.22 mag), and that in five of the six cases the index $n$ is close to 
unity. This result confirms that the systems 
are disk-dominated\footnote{ We note that disk-like systems at $z\sim$2 should not
necessarily be thought of as spiral galaxies. Disks mostly refer to all smooth or 
somewhat clumpy light distributions that follow an exponential profile. }.
However, this analysis needs to be performed on large samples to investigate how 
frequent or efficient the formation of bulges by ULIRGs is at $z$=2.

A similar bulge-to-disk decomposition was performed by \cite{zheng} on lower 
redshift ($z\lesssim$1.2) LIRGs and ULIRGs. Their sources were detected at 15 
\micron\ using ISO Camera data in the Canada-France Redshift Survey (\citealt{flores}). 
Their sources were observed with the {\it HST} Wide Field Planetary Camera 2 in the 
$B$,$V$, and $I$ bands. \cite{zheng} also find that the six single-nucleus ULIRGs
in their sample likewise have small or negligible bulges. \cite{melbourne} performed
a morphological analysis of IR-bright sources up to $z$=1 in the Great Observatories
Origins Deep Survey (GOODS) 24 \micron\ image (R. R. Chary et al.~2008, in preparation) and 
found that $z$=1 LIRGs are mostly spirals or peculiar. In the combined GOODS and 
Galaxy Evolution from Morphology and SED (GEMS) fields, \cite{bell05} find that 
approximately 50\% of the sources are spiral galaxies at $z$=0.7.

\subsection{Rest-frame Optical vs MIR Galaxy Properties} 

\begin{figure*}
\centering
\includegraphics[width=16cm]{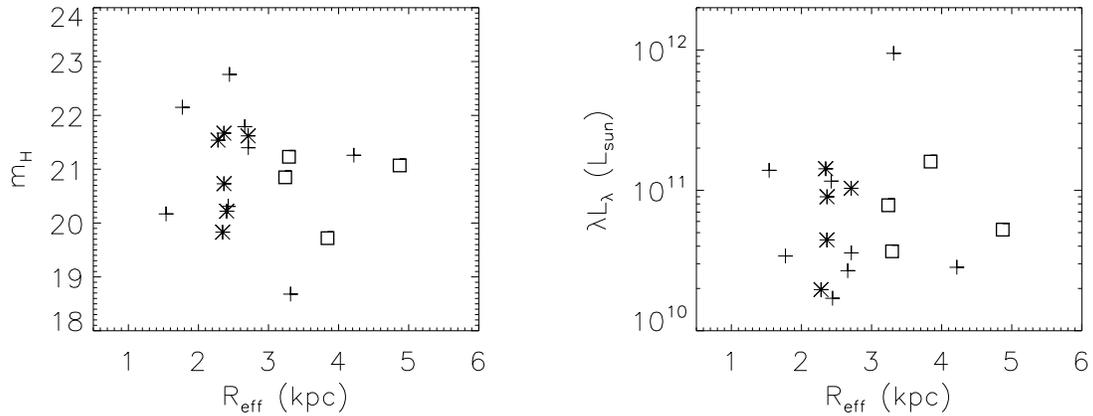}
\caption{\label{fig:mir} Observed $H$-band (rest-frame optical) photometric 
properties as a function of the MIR spectral type. The apparent magnitudes 
({\it left}) and the luminosities ({\it right}) are plotted vs. the effective 
radii for AGNs (stars), starbursts (boxes) and obscured systems (crosses).
In this diagram, we are not using sources with composite IRS spectra, i.e.,
sources with simultaneous PAH emission and SiO absorption features, 
and AGN continua (see Table \ref{tab:obs}). 
} 
\end{figure*}

In Figure~\ref{fig:mir} we examine the dependence of the rest-frame optical 
morphological parameters of the sources in our sample on their MIR spectral
classification. The optical magnitudes and luminosities are plotted as a function 
of the effective radii for the sources with well-defined MIR spectral types (see 
Table~\ref{tab:obs}). Both the magnitudes and the luminosities are statistically 
indistinguishable for all types of sources. However, the various systems segregate 
along the effective radius axis. AGN-dominated sources are compact systems with 
\reff\ of $\sim$2 kpc, whereas PAH-dominated, starburst sources are as extended
as 5 kpc. The obscured sources have a large range of effective radii, comprising
both compact and extended systems.

\begin{figure*}
\centering
\includegraphics[width=16cm]{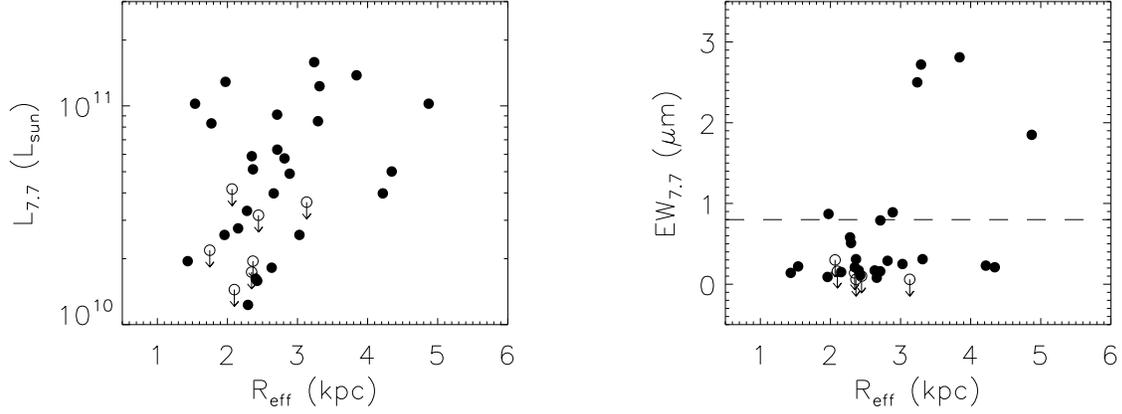}
\caption{\label{fig:pah} Luminosity ({\it left}) and equivalent width ({\it right}) 
of the 7.7 \micron\ PAH feature as a function of the effective radius.
All sources with 7.7 \micron\ feature detections are used in this diagram,
regardless of the dominant component of the MIR spectrum. Filled circles
correspond to sources with detections and open circles correspond to sources
with upper limits on their 7.7 \micron\ luminosities. The dashed line
on the right panel corresponds to the threshold that \cite{sajina07} set to
discriminate between sources with strong and weak PAH emission.
}
\end{figure*}

In Figure~\ref{fig:pah} we plot the 7.7 \micron\ PAH feature luminosity, $L_{7.7}$, 
and equivalent width, EW$_{7.7}$ (\citealt{sajina07}), as a function of the 
rest-frame optical effective 
radius. In this plot, we do not distinguish between sources of different MIR spectral 
types. We find that high PAH luminosities appear in both compact and extended sources 
but the lowest values of $L_{7.7}$ appear in compact systems. The 7.7 \micron\ 
equivalent width does not correlate with the optical extent of the system, but 
the sources with the highest EWs are extended. For compact objects, this result 
probably reflects the dilution of the PAH features by the AGN continuum, since
compact sources have on average a stronger AGN component in both IR and optical 
wavelengths than most extended systems do (\citealt{tran01}; \citealt{veilleux02}). 
This result could be a confirmation of a scenario in which ULIRGs with 
warm MIR/FIR SEDs are more AGN-dominated than ULIRGs with cold MIR/FIR SEDs 
(\citealt{sanders88}). Still, compact sources with high PAH luminosity exist. 

\begin{figure*}
\centering
\includegraphics[width=16cm]{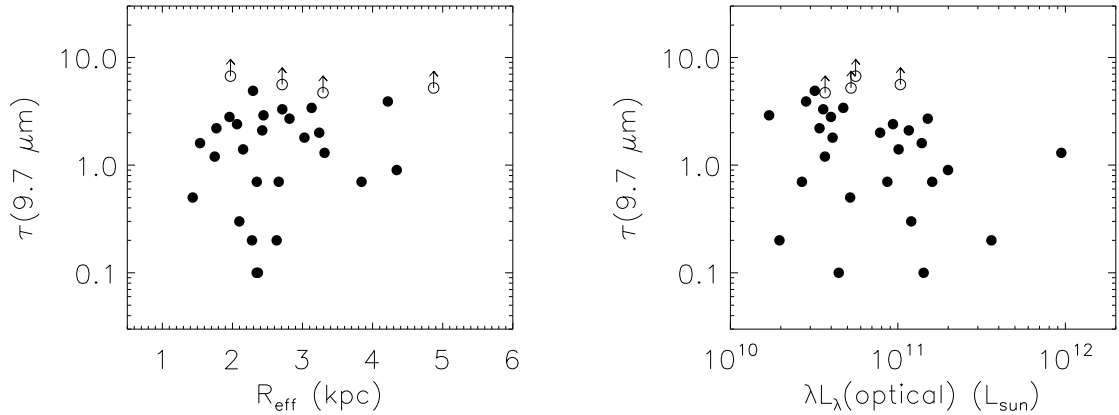}
\caption{\label{fig:tau} MIR optical depth, as measured from the 9.7 \micron\ 
SiO feature, is plotted as a function of the rest-frame optical effective radius
({\it left}) and the rest-frame optical luminosity ({\it right}). Filled circles
are sources with detections and open circles are sources with lower limits on 
their MIR obscuration.}
\end{figure*}

In Figure~\ref{fig:tau}, we plot the MIR optical depth, $\tau$,  as a function
of the rest-frame optical effective radius and luminosity of each source. The
value of $\tau$ is derived from the SiO absorption feature at 9.7 \micron\ 
(\citealt{sajina07}). The spatial distribution of the silicates that are responsible for the 
MIR obscuration is not correlated with the rest-frame optical stellar light profile. The 
optical depth of the MIR obscuring medium does not correlate with the stellar
luminosity either:  obscured systems in the MIR can be bright in the optical 
wavelengths (e.g., MIPS~8196). This result could imply that the MIR obscuring medium 
is circumnuclear  (\citealt{spoon}; \citealt{sajina07}).


\section{Discussion: Trigger mechanisms of z$\sim$2 vs local ULIRGs}
\label{sec:comp}

A qualitative morphological classification of the ULIRGs in our sample indicates
that there is a significant fraction of sources that appear to be quiescent
or compact. The number of sources that have no prominent signs of interaction is 
48\%. This fraction is high in comparison with the local ULIRG samples, in which
only 2\% of the sources are not distorted (\citealt{kim02}; \citealt{veilleux06}).  
The high fraction of quiescent sources could be attributed to flux or surface 
brightness limitations of our observations (see \S\ref{sec:host}). However, 
{\it HST} NICMOS, Advanced Camera for Surveys (ACS), and Space Telescope Imaging 
Spectrograph (STIS) observations of 1$\le z\le$3 submillimeter galaxies (\citealt{chapman}; 
\citealt{swinbank}) indicate that $\sim$80\% of these sources have signs of interaction. 
Although color effects may change this fraction, it is possible that the differences
could alternatively be specific to the selection criteria of each sample. 
Since our objects are brighter than 0.9 mJy at 24 \micron, they have warmer MIR/FIR 
SEDs than submillimeter galaxies (Sajina et al.~2008). According to a scenario in which 
a warm ULIRG phase follows a cold ULIRG phase in the same evolutionary sequence (\citealt{sanders88}), 
our sources appear to be less perturbed than submillimeter galaxies because they could be closer to 
dynamical relaxation. In this picture, the submillimeter galaxies could be the cold, binary 
mergers and the 24 \micron -selected ULIRGs could be their warm, merged counterparts. 
This possibility is supported by the findings of \cite{bridge}, who showed that the 
likelihood of a merger to be in a close pair is 5 times higher for sources brighter 
than 0.1 mJy at 24 \micron\ than for sources below this limit. 

To investigate for (dis)similarities in the properties of z$\sim$2 and local ULIRGs, 
we compared the number of on-going major mergers\footnote{
We note that all comparisons between the local and the z$\sim$2 ULIRGs presented in this
Section are based on the rest-frame optical luminosity ratios of the interacting galaxies. 
Since this ratio could be affected by extinction, it could deviate from its intrinsic value. 
Local ULIRGs that appear to be minor mergers in optical images but major mergers
in near-IR images and vice versa exist (\citealt{kim02}; \citealt{surace00b}). The strength
of this effect can vary significantly from source to source, depending on the 
geometric distribution of its obscuring medium.}
in our sample and in the entire 1 Jy catalog of mainly local ULIRGs (\citealt{kim02}). 
Of the 118 sources in the 1 Jy catalog, 40 (or 34\%) are binaries with an optical 
($B-$ to $R-$ band) luminosity ratio $\le$3:1, or $\Delta m$$\le$1.2, and 
1.5$\le$$d_n$$\le$30 kpc\footnote{
We chose our fiducial nuclear separation threshold to be 30 kpc since there is only 
one binary system in the 1 Jy sample with $d_n$ that exceeds this value.}
(\citealt{kim02}; \citealt{surace00b}). In 
the z$\sim$2 sample, 2 (or 6\%) of the 33 systems (MIPS~16059 and MIPS~22530) are 
currently interacting binaries with $\Delta m$$ \le$1.2 and 1.5$\le$$d_n$$\le $30 kpc. We 
ran a Monte Carlo code to quantify the significance in the difference of these fractions. 
We created a parent population of 118000 sources, 40000 of which are binaries  that follow 
the $d_n$ and $\Delta m$ distribution of the 1 Jy sample binaries. We randomly chose 
33 sources and computed the probability $p$ that a number $n_S$ of the 33 sources meet the 
criteria 1.5$\le$$d_n$$\le$30 kpc and $\Delta m$$\le$1.2. We iterated the procedure for $10^5$ 
times and found that $p$ is insignificant (0.014\%) for $n_s$$=$2. Therefore, our sample statistically 
differs from the entire 1 Jy sample. Four more sources in our sample (MIPS~464, MIPS~15840, MIPS~15880, 
and MIPS~22651) have neighbours that satisfy our nuclear separation and luminosity ratio 
criteria. However, they have no signs of interaction with their neighbours. If we assumed 
all these sources to be interacting, the maximum fraction of major-merger pairs in our 
sample would increase to 18\%. In that case, the probability $p$ would be 3.0\%, which 
is significant but still small.

We then examined whether the sources in our sample are the analogs of the warmest 
local ULIRGs by comparing the number of on-going major mergers in both our sample and 
in a local luminosity-matched control sample. We selected the control sample as the 1 
Jy catalog sources that have $\lir \ge 10^{12.5}$\lsun\ (\citealt{kim02}) for the
cosmological parameters adopted in this paper. The local control sample has 16 sources 
that typically have warm MIR SEDs (e.g., \citealt{tran01}; \citealt{armus}). Two or 
13\% of these sources are binaries with $\Delta m$$\le$1.2 and 1.5$\le$$d_n$$\le $30 kpc. 
This fraction of binary systems is statistically indistinguishable from that in our 
sample (6\% $-$18\%). Therefore, the z$\sim$2 ULIRGs morphologically resemble their warmest 
local analogs. However, they differ from their local analogs in their MIR spectral properties. 
IRS spectroscopy indicates that the majority of the most IR luminous local sources have a strong 
AGN component in the MIR (L. Armus et al. 2008, in preparation), unlike the $z$$\sim$2 sample in
which systems with strong
starburst-related features exist. One such example is MIPS 22530, which has a MIR spectrum with
strong PAH emission and \lir = $8 \times 10^{12}$ \lsun. We conclude that there is no local ULIRG
\hbox{(sub)sample} that is likely to comprise sources with simultaneous compact morphologies, high IR 
luminosities, and strong starburst-related features, which are characteristic of these $z$$\sim$2 ULIRGs.  

In the local Universe, minor mergers are associated with LIRGs instead of ULIRGs, 
although few atypical examples of ULIRGs that are triggered by such mergers exist, 
e.g., IRAS 12127-1412 (\citealt{kim02}). Up to one-fifth of the ULIRGs in our sample 
could be associated with on-going minor mergers. Six sources (MIPS~78, MIPS~180, 
MIPS~8242, MIPS~8342, MIPS~8493, and MIPS~16113) have low surface brightness 
neighbours within a radius of 10 kpc and could be close interacting binaries. 
We used galaxy number counts $N_g$ per square degree and magnitude to find the probability 
of randomly having a faint background or foreground source within a projected distance 
of 10 kpc or 1\farcs2 at $z$=2. We constructed a Monte Carlo simulation to randomly 
distribute $N_g$ galaxies over 1 deg$^2$ and to compute the projected separation 
of each source from its nearest neighbour. For $N_g$=150000 galaxies in the range 
$m_H$=22.5 to 23.5 (\citealt{metcalfe06}), we found that 5 out of every 100 sources 
have a neighbour within the given separation due to projection effects. We then 
selected 33 sources from this parent population and computed the probability that at
least six galaxies have neighbours due to projection effects. We iterated the procedure 
for $10^5$ times and found that the probability is 0.52\%, which indicates that some 
of these sources are likely involved in on-going minor mergers. 

In \S\ref{sec:host}, we showed that five of the six brightest 
sources in our sample have a small S\'ersic index ($<1.35$) and bulge-to-disk ratio 
($\le 0.57$). Such systems can evolve to disk galaxies with small bulges, unlike local 
ULIRGs. Regardless of how representative this result is for the majority of 
$z$=2 ULIRGs, it provides further evidence that the dynamical configuration of IR-bright 
sources can differ at $z$$\sim$2 and in the local Universe.

A plausible explanation regarding why some sources differ from 
their local analogs is that the $z$$\sim$2 systems, being irregular and disky, 
could have high fractions of gas and dust. Simulations have shown that for high gas 
fractions, the outcome of a merger is sensitive to the treatment of the interstellar 
medium and could lead to the formation of rotationally supported remnants 
(e.g., \citealt{robertson}). Another scenario is that the strong star-formation
episodes occur later in the merger timescales of $z$$=$2 ULIRGs than those of 
local ULIRGs due to differences in the dynamical configuration of their progenitors.


\section{Conclusions}
\label{sec:conc}
We acquired \hst\ NIC2 images of 33 $z\sim$2 ULIRGs to investigate the 
dynamical triggers of high-$z$, 24 \micron -selected populations, revealed
by \spi . Having a resolution element of 1.3 kpc at $z$=2, our $H$-band images 
enable us to perform a comprehensive morphological analysis of these sources. Our 
findings are summarized as follows.
\begin{enumerate}
\item
The rest-frame optical luminosities of the sources in our sample are of the order 
$10^{10}-10^{11}$ \lsun\ and their mean half-light radius equals 2.66 kpc. On
average, this observed \reff\ value is half of that of local ULIRGs, but surface
brightness dimming might account for all or part of this difference.
\item
Taking into account binary systems, tidal tails and other irregular morphologies, the 
number of sources apparently involved in interactions is 17 out of 33. This is a lower limit 
since unresolved binaries, possible faint companions, or diluted low surface-brightness 
features (of $\mu_H$$>$22 mag arcsec$^{-2}$) could exist in our sample. We simulated the 
effects of surface brightness dimming on our photometric analysis and found that extended 
structures such as tidal tails can typically be seen in systems as faint as $m_H$=20. The mean 
apparent magnitude of the sources in our sample equals $m_H$=20.89. 
\item
Up to one-fifth of the sources in our sample could be associated with minor mergers. The 
number of binary systems with components of comparable luminosity and signs of on-going 
interaction is only 2. This major-merger pair fraction is significantly smaller than that 
of the local ULIRG population. 
\item
We performed a bulge-to-disk decomposition for all the sources that are brighter 
than $m_H$=20. We found that these sources are disk-like with small or negligible bulges.   
\item
The AGN-dominated systems have the most compact and the PAH-dominated systems have the most 
extended morphology in the rest-frame optical images. The luminosity of the 7.7 \micron\ PAH feature 
is often, but not always, suppressed in the compact systems. Its EW is high only in extended sources. 
The MIR obscuration correlates neither with the radial extent nor with the rest-frame optical luminosity 
of the sources. 
\end{enumerate}
We conclude that a nonnegligible fraction of the sources in our sample do not resemble local 
ULIRGs, which are produced by major mergers and form elliptical galaxies. However, it is 
unclear whether this result is characteristic of most $z$=2 ULIRGs or whether it 
relates to our selection criteria. In either case, it provides evidence that several 
interaction types can be associated with a ULIRG at z$=$2.


\acknowledgments

K. D. wishes to thank Lee Armus and Lisa Storrie-Lombardi for useful discussions. This
work is based on observations made with the NASA/ESA {\it Hubble Space Telescope}, obtained 
from the data archive at the Space Telescope Science Institute. STScI is operated by the 
Association of Universities for Research in Astronomy, Inc. under NASA contract NAS 5-26555.
It is also based in part on observations made with the Spitzer Space Telescope, 
which is operated by the Jet Propulsion Laboratory, California Institute of Technology 
under a contract with NASA. Support for this work was provided by NASA through an award 
issued by JPL/Caltech. 

\clearpage


\appendix
\section{APPENDIX: Surface Brightness Dimming Simulations}
\label{sec:dimming}

A question that naturally arises during the photometric analysis of intermediate 
and high $z$ sources is what the effects are of surface brightness dimming
on the analysis performed. Although this issue has been addressed several times 
(e.g., \citealt{giavalisco}; \citealt{hibbard}), the results 
significantly depend on the type of sources under examination and the setup 
(e.g., central wavelength) of the observations. We therefore constructed a model 
to specifically address the following questions: (1) Up to which magnitude can 
secondary, faint nuclei be observed? (2) How do the observed magnitudes and
effective radii relate to their intrinsic values? And (3) at which point does 
the bulge-to-disk decomposition become unreliable?

Our surface-brightness-dimming simulations use templates of local galaxies that 
are shifted to $z$=2. The templates were selected in a manner such that a large 
number of galaxy types are examined. We used archival data (PIs: A. S. Evans, R. Sharples) 
of two ULIRGs, one LIRG, one quasar, and one elliptical galaxy in the local 
Universe  observed with ACS. 
Namely, these sources are IRAS~23128-5919, which is a binary ULIRG separated 
by 4.3 kpc (\citealt{dasyra06a}), IC 4687, which is a LIRG with a significantly 
distorted disk, IRAS~09111-1007, which has secondary components, NGC~3156, and 
PG~1351+640. The sources were observed using either the 435W, the 475W, or the 625W 
filter, similar to the rest-frame wavelengths of our observations. 

To simulate $z$=2 galaxies from local Universe templates, we followed the 
prescriptions of \cite{giavalisco}. First, we scaled down the angular size 
$\theta$ of the template galaxy. The binning factor $b$ was given by
{\large
\begin{equation}
\hbox{$ b = \frac{\theta _{z_f} s_{z_i}}{\theta _{z_i} s_{z_f}}$,}
\end{equation}
}
where $z_i$ is the initial redshift of each template, $z_f$ is the final redshift
for all templates (which is equal to 2), and $s_z$ is the pixel scale of the 
instrument used to perform the observations at a given redshift $z$. In our case, 
the instrumental pixel scales correspond to  
$s_{z_f}$=$s({\rm NICMOS})$=$0\farcs076$ and $s_{z_i}$=$s({\rm ACS})$=$0\farcs048$.
To ensure that the physical size of each source remains the same at 
all redshifts, its angular size ratio at $z_f$ and $z_i$ needs to be 
{\large
\begin{equation}
\hbox{$\frac{\theta _{z_f}}{\theta _{z_i}}=\frac{(1+z_f)^2 D_L(z_i)}{(1+z_i)^2 D_L(z_f)}$,}
\end{equation}
}
where $D_L(z)$ is the luminosity distance of a source at redshift $z$.

We then accounted for the surface brightness dimming by multiplying the rescaled images
with a dimming factor $d$ that was equal to
{\large
\begin{equation}
\hbox{$d=x \frac{\alpha_{z_i}}{\alpha_{z_f}} \left( \frac{s_{z_f}}{s_{z_i}}\right) ^2 \frac{\Delta \lambda_{z_f}}{\Delta \lambda_{z_i}} \left( 
\frac{1+z_i}{1+z_f} \right) ^5$.}
\end{equation}
}
The factor $\alpha_{z}$ is the constant that converts counts (or electrons) per 
second to flux. For the 160W filter of NICMOS, $\alpha_{z_f}$ is 
$1.75\times10^{-19}$ ergs cm$^{-2}$ $\ang ^{-1}$ count$^{-1}$. For ACS 435W, 475W, and 
625W, the factor $\alpha_{z_i}$ is equal to 3.14, 1.81, and 1.20 $\times10^{-19}$ ergs 
cm$^{-2}$ $\ang^{-1}$ electron$^{-1}$, respectively.
The quantity $\Delta \lambda_{z}$ is the filter width for the instrument used for the
observations at redshift $z$; it is equal to 4000, 1038, 1458, and 1442 \ang\ for NICMOS 
160W, ACS F435W, ACS, F475W, and ACS F625W, respectively. The factor $x$ 
is used to match the intrinsic luminosity of the template with that of a source of 
a given magnitude at $z$=2. Given the observed magnitudes of our sources, we set the 
initial $m_H({\rm init})$ value to either 19, 20, 21, or 22 mag. 

\begin{figure*}
\centering
\includegraphics[width=16cm]{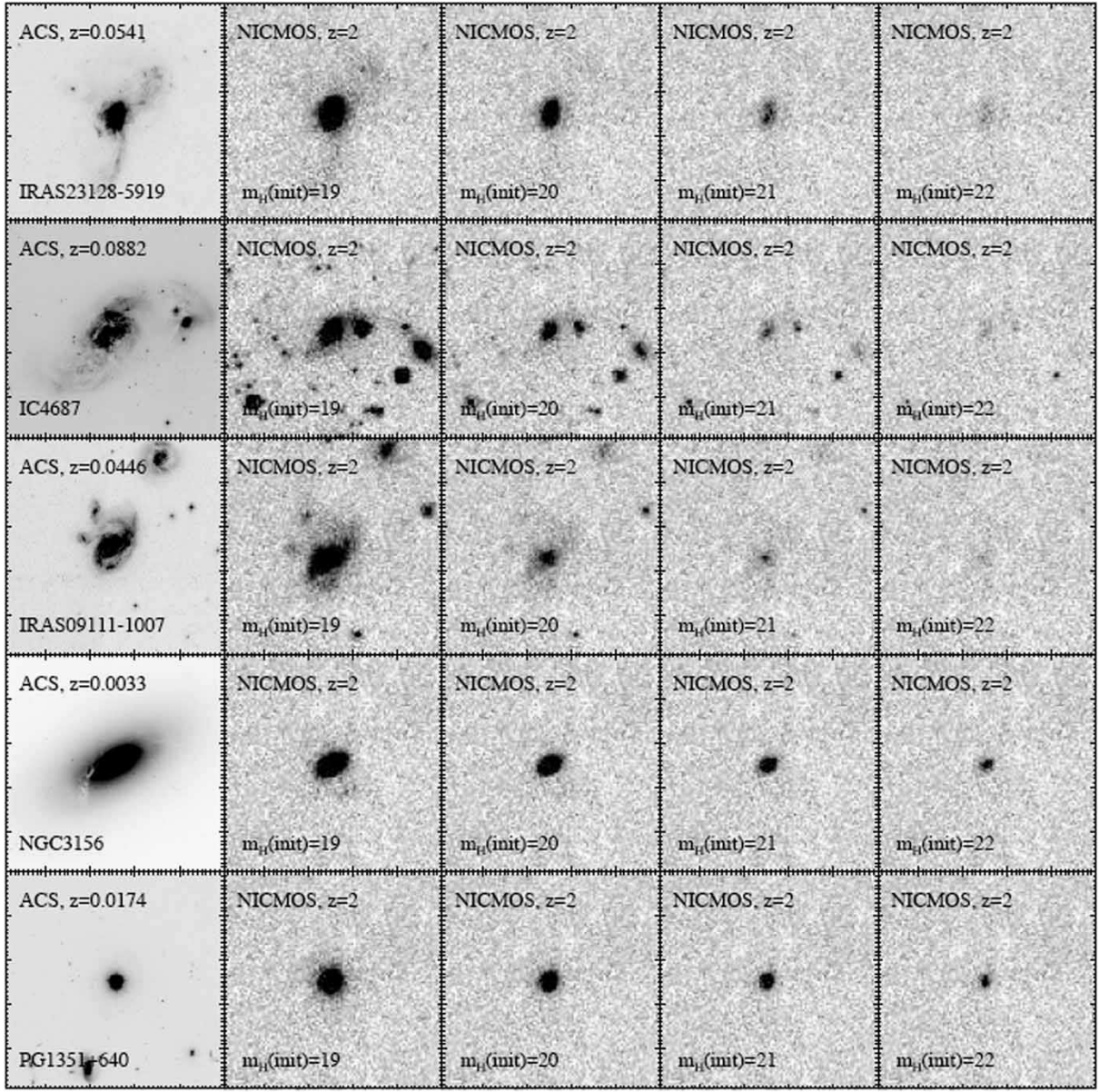}
\caption{\label{fig:simulations} Results from a surface-brightness dimming
model that simulates how local galaxies of different morphological types would 
be observed at $z$=2. The first column shows the 40$'' \times$ 40$''$ 
ACS images of the local template galaxies. The next four columns show the 
simulated NIC2 images of the same galaxies that are scaled to match the 
intrinsic luminosity of a $z$=2 source with an apparent magnitude of 
$m_H({\rm init})$=19, 20, 21, and 22 respectively, prior to its dilution 
into the sky background. The magnitudes of these objects as observed after 
the addition of the background are tabulated in Table~\ref{tab:simulations}.
The size of the NIC2 simulated images corresponds to $3\farcs 8 
\times 3\farcs 8$, similar to the observations in Fig.~\ref{fig:images}.} 
\end{figure*}

After convolving the resulting images with the NIC2 PSF, we added background noise to 
them. For this step, we computed and added to our images the median sky image of 
all data sets. To imitate the reduction of the real data, we then subtracted from these 
images the sky background of a random data set. We present the final simulated images 
in Figure~\ref{fig:simulations}. We processed these images with {\it sextractor} and 
{\it galfit} in a manner similar to the NICMOS data. The magnitudes and effective
radii, as they would be observed following the addition of the sky background, are 
denoted as $m_H({\rm obs})$ and $\reff ({\rm obs})$ and are presented in 
Table~\ref{tab:simulations}.

Already a visual inspection of the simulation results indicates that sources of
$m_H({\rm init})$=19 preserve to a large extent their initial light profiles. The 
sources of $m_H({\rm init})$=20 lose part of their low surface brightness 
components but still maintain an extended structure. Mergers of equal luminosity 
systems can be identified all the way up to $m_H({\rm init})$=21 or an observed
magnitude $m_H({\rm obs})$ that is $\lesssim 22.0$. However, faint 
secondary nuclei are mostly seen up to $m_H({\rm init})$=20. For sources with 
$m_H({\rm init})\ge$21, one could often be unable to disentangle between minor 
mergers, single-nucleus remnants of major mergers, or quiescent sources. 

To quantify the surface-brightness-dimming effects on the photometric properties
of our sources, we ran {\it sextractor} for all simulated galaxies. The results 
are presented in Table~\ref{tab:simulations} and depend on the initial light 
profile distribution. Compact distributions, such as PSFs and bulge components 
are more resilient to the surface brightness dimming than extended, disk-like 
distributions. For instance, we found that a spiral or irregular galaxy of 
$m_H({\rm init})=$19 and 20 loses about 27\% and 38\% of its light, respectively, 
because of dimming effects. The observed effective radius is smaller by less 
than 14\% and 25\%, respectively, than its initial value. At $m_H({\rm init})=$21,
about half of the light is lost, leading to \reff\ values that are underestimated 
by a factor of 2. Systems of $m_H({\rm init}) \ge$22 are typically undetected at 
3$\sigma$ levels.
In contrast, the light losses are small for an elliptical galaxy. Its flux 
is decreased by 1\% up to 22\% and its radius is reduced by 2\% up to 17\% from 
its intrinsic values at $m_H({\rm init})=$19 and 22. 

We then aimed to address the reliability of the bulge-to-disk decomposition technique by 
examining how different its results are when shifting a local source to $z$$=$2, and 
when increasing (or decreasing) the brightness of a $z$$=$2 system by 1 mag.
For this purpose, we ran {\it galfit} for some local sources and their simulated 
$z$$=$2 images, namely, the disk-dominated main component of IC~4687 and the elliptical 
galaxy NGC~3156. 
IRAS~23128-5919 and IRAS~09111-1007 are too perturbed to allow a decomposition. Compact 
light profiles such as those of PG 1351+640 were not analyzed either because they
are degenerate between a PSF, and a small, unresolved bulge or disk. Since the light
profiles of IC~4687 and NGC~3156 are close to those of a disk and a bulge, we examined the 
value of their S\'ersic index as a function of redshift and integrated magnitude. We find 
that the $z$$=$2 images of IC~4687 have $n=$1.05 and 1.06, respectively, for $m_H({\rm init})$=19 
and 20. For comparison, the actual S\'ersic index of IC~4687 at a redshift of 0.0173 
is 1.20. The $z$$=$2 images of NGC~3156 have $n=$5.67 and 5.28 for $m_H({\rm init})$=19 
and 20, whereas at $z$$=$0.004396, NGC~3156 has $n=$6.70. The small differences for 
the elliptical galaxy are likely related to the spatial dilution of the initial (ACS) 
PSF with redshift. This comparison shows that the shape of the light profile can be accurately 
predicted at $z$$=$2. However, we wish to note that in these simulations, we used different 
magnitudes but identical initial structural parameters for the $z$$=$2, $m_H({\rm init})$=19 
and 20 images of each source to properly examine the effects of surface brightness dimming. 
For observational data there is an extra difficulty; the best {\it galfit} 
solution can significantly depend on the initial parameter values. For the systems of 
$m_H({\rm init})$=19, a reasonable input file can be constructed by visual 
inspection of the images, e.g., by determining the size of the extended structures. 
For the systems of $m_H({\rm init})\ge$20, the surface brightness dimming does 
not always allow for the determination of an optimal input file. {\it Galfit} 
can then fall into a local (instead of the global) $\chi^2$ minimum. To avoid
degenerate solutions, we opt to apply this decomposition technique only to sources 
that are brighter than $m_H$=20.


\clearpage


\begin{deluxetable}{ccccccccc}
\tablecolumns{9}
\tabletypesize{\tiny}
\tablewidth{0pt}
\tablecaption{\label{tab:obs} Source List and Observation Log}
\tablehead{
\colhead{Galaxy} 
& \colhead{RA \tablenotemark{a}} & \colhead{Dec \tablenotemark{a}} 
& \colhead{$z$ \tablenotemark{b}} 
& \colhead{log($L_{14}$/\lsun) \tablenotemark{c}} 
& \colhead{log(\lir/\lsun) \tablenotemark{c}} 
& \colhead{MIR \tablenotemark{b}}
& \colhead{t$_{\rm int}$ \tablenotemark{d}} 
& \colhead{$\Delta \lambda_{\rm rest}$ \tablenotemark{e}} \\

\colhead{} & \colhead{(2000)} & \colhead{(2000)} & \colhead{} & \colhead{}
& \colhead{} & \colhead{type} & \colhead{(s)} & \colhead{(\ang )}
}

\startdata
MIPS42 & 17:17:58.54 & +59:28:16.18 & 1.95  & 12.68 & \nodata & obscured & 5120 & 4750$-$6100 \\
MIPS78 & 17:15:38.18 & +59:25:40.08  & 2.65 & 12.76 & \nodata  & mixed & 5248 & 3840$-$4930 \\
MIPS180 & 17:15:43.73 & +58:35:32.97 & 2.47 & 12.57 & \nodata & obscured & 5248 & 4040$-$5190 \\
MIPS227 & 17:14:55.99 & +58:38:16.44 & 1.87 & 12.39 & \nodata & mixed & 2560 & 4880$-$6270 \\
MIPS289 & 17:13:50.04 & +58:56:54.45 & 1.86 & 11.80 & 12.9    & PAH & 2560 & 4900$-$6300 \\
MIPS464 & 17:14:39.60 & +58:56:32.07 & 1.85 & 12.16 & \nodata & mixed & 5120 & 4910$-$6320 \\
MIPS506 & 17:11:38.52 & +58:38:38.58 & 2.52 & 12.74 & \nodata & mixed & 2560 & 3980$-$5110 \\
MIPS8196 & 17:15:10.17 & +60:09:54.54 & 2.60 & 12.53 &\nodata  & obscured & 2688 & 3890$-$5000 \\
MIPS8242 & 17:14:33.19 & +59:39:11.31 & 2.45 & 12.74 & \nodata & mixed & 2560 & 4060$-$5220 \\
MIPS8245 & 17:15:36.31 & +59:36:14.73 & 2.70 & 12.55 & \nodata  & mixed & 5248 & 3780$-$4870 \\
MIPS8327 & 17:15:35.62 & +60:28:24.49 & 2.48 & 12.40 & \nodata & mixed & 5248  & 4020$-$5170 \\
MIPS8342 & 17:14:11.45 & +60:11:09.20 & 1.57 & 11.93 & 12.7  & AGN & 2688 & 5450$-$7000\\
MIPS8493 & 17:18:04.99 & +60:08:32.28 & 1.80 & 11.49 & \nodata & PAH & 2688 & 5000$-$6430 \\
MIPS15840 & 17:19:22.44 & +60:05:00.42 & 2.30 & 12.50 & \nodata  & AGN & 5248 & 4240$-$5450 \\
MIPS15880 & 17:21:19.61 & +59:58:17.87 & 1.64 & 12.51 & 12.9 & obscured & 2560 & 5310$-$6820 \\
MIPS15928 & 17:19:17.47 & +60:15:19.62 & 1.52 & 12.04 & \nodata & mixed & 2688 & 5560$-$7140 \\
MIPS15949 & 17:21:09.21 & +60:15:01.66 & 2.15 & 12.45 & \nodata & AGN & 2688 & 4440$-$5710 \\ 
MIPS15958 & 17:23:24.21 & +59:24:55.62 & 1.97 & 12.28 & \nodata & mixed & 2560 & 4710$-$6060 \\ 
MIPS15977 & 17:18:55.66 & +59:45:45.61 & 1.85 & 12.28 & 13.0 & AGN  & 2560 & 4910$-$6320 \\
MIPS16059 & 17:24:28.34 & +60:15:33.08 & 2.43 & 12.46 & \nodata & mixed & 5376 & 4080$-$5250 \\
MIPS16080 & 17:18:44.83 & +60:01:15.82 & 2.04 & 12.18 & \nodata & obscured & 2688 & 4610$-$5920 \\
MIPS16095 & 17:23:59.77 & +59:57:52.42 & 1.81 & 12.12 & \nodata & mixed & 5120 & 4980$-$6410 \\
MIPS16113 & 17:21:26.47 & +60:16:46.27 & 1.90 & 11.88 & \nodata & obscured & 2688 & 4830$-$6210 \\
MIPS16122 & 17:20:51.50 & +60:01:48.75 & 1.97 & 12.13 & \nodata & mixed & 5376 & 4710$-$6060 \\
MIPS16144 & 17:24:22.03 & +59:31:50.63 & 2.13 & 12.01 & \nodata & PAH & 2560 & 4470$-$5750 \\
MIPS22204 & 17:18:44.40 & +59:20:00.96 & 2.08 & 12.63 & \nodata & obscured & 2560 & 4550$-$5850 \\
MIPS22277 & 17:18:26.73 & +58:42:42.12 & 1.77 & 12.36 & \nodata & mixed & 2560 & 5050$-$6500 \\ 
MIPS22303 & 17:18:48.87 & +58:51:14.87 & 2.34 & 12.57 & \nodata & obscured & 5248 & 4190$-$5390 \\
MIPS22530 & 17:23:03.31 & +59:16:00.55 & 1.96 & 12.10 & 13.0 &    PAH & 2560 & 4730$-$6080 \\
MIPS22558 & 17:20:45.15 & +58:52:21.50 & 3.20 & 12.88 & \nodata & AGN & 5248 & 3330$-$4290 \\
MIPS22651 & 17:19:26.54 & +59:09:29.12 & 1.73 & 11.86 & 12.9 & mixed & 2560 & 5130$-$6590 \\
MIPS22661 & 17:18:19.63 & +59:02:42.68 & 1.75 & 11.93 & \nodata & AGN & 5248 & 5090$-$6550 \\
MIPS22699 & 17:20:47.52 & +59:08:14.71 & 2.59 & 12.17 & \nodata  & mixed & 5248 & 3890$-$5010 \\
\enddata

\tablenotetext{a}
{Units of right ascension are hours, minutes and seconds, and units of 
declination are degrees, arcminutes, and arcseconds. The coordinates of
the sources are those of the IRAC 3.6 \micron\ catalog of \cite{lacy05}.}

\tablenotetext{b}
{The redshifts and the MIR spectra used for the type classifications are from 
\cite{sajina07}. The spectral type nomenclature adopted here simply denotes the
dominant component of the MIR spectrum, namely the AGN continuum, the PAH emission
features, or the SiO absorption feature at 9.7 \micron. If the dominant component 
is not obvious, we simply describe it as a mixed system.}

\tablenotetext{c}
{The IR and 14 \micron\ luminosities are from Sajina et al. (2008)
and \cite{sajina07} respectively. The 14 \micron\ luminosity can be considered
a lower limit of \lir\ when the latter is not available.}

\tablenotetext{d}
{Total integration time of the NIC2 exposures.}

\tablenotetext{e}
{Rest-frame wavelength range of the observations performed with the NIC2 camera and the F160W filter.}

\end{deluxetable}

\begin{deluxetable}{ccccccc}
\tablecolumns{7}
\tabletypesize{\tiny}
\tablewidth{0pt}
\tablecaption{\label{tab:phot} Global Photometric Parameters}
\tablehead{
\colhead{Galaxy} & \colhead{$m_H$\tablenotemark{a}}
& \colhead{$M_H$ \tablenotemark{a}}   
& \colhead{$<\mu_H >$ \tablenotemark{a}}  
& \colhead{$\reff$ \tablenotemark{a}}  
& \colhead{$\nu {L_{\nu}}$ (opt) \tablenotemark{a}}  
& \colhead{interaction} \\

& \colhead{} & \colhead{} & \colhead{(mag arcsec$^{-2}$)} & \colhead{(kpc)}& \colhead{(\lsun)}& \colhead{signs}
}

\startdata
MIPS42 & 21.79 & -24.10 & 21.29 & 2.66 & 2.68$\times{10^{10}}$ & none \\
MIPS78 & 21.98 & -24.72 & 21.95 & 3.13 & 4.72$\times{10^{10}}$ & none\tablenotemark{b} \\
MIPS180 & 22.15 & -24.37 & 20.85 & 1.77 & 3.41$\times{10^{10}}$ & none\tablenotemark{b} \\
MIPS227 & 18.85 & -26.93 & 18.32 & 2.63 & 3.62$\times{10^{11}}$ & none \\
MIPS289 & 19.72 & -26.04 & 20.01 & 3.84 & 1.60$\times{10^{11}}$ & distorted\tablenotemark{c} \\
MIPS464 & 21.45 & -24.30 & 20.62 & 2.29 & 3.19$\times{10^{10}}$ & distorted\tablenotemark{d} \\
MIPS506 & 21.66 & -24.91 & 20.60 & 1.97 & 5.60$\times{10^{10}}$ & distorted\\
MIPS8196 & 18.68 & -27.97 & 18.76 & 3.31 & 9.46$\times{10^{11}}$ & distorted \\
MIPS8242 & 20.21 & -26.28 & 20.85 & 4.34 & 1.99$\times{10^{11}}$ & distorted\tablenotemark{e} \\
MIPS8245 & 22.21 & -24.54 & 21.17 & 1.96 & 3.98$\times{10^{10}}$ & none  \\
MIPS8327 & 21.06 & -25.47 & 20.09 & 2.07 & 9.35$\times{10^{10}}$ & distorted\tablenotemark{c} \\
MIPS8342 & 21.54 & -23.77 & 20.68 & 2.28 & 1.96$\times{10^{10}}$ & none\tablenotemark{b} \\
MIPS8493 & 21.23 & -24.45 & 21.18 & 3.29 & 3.68$\times{10^{10}}$ & distorted\tablenotemark{e} \\
MIPS15840 & 21.67 & -24.66 & 20.96 & 2.36 & 4.44$\times{10^{10}}$ & none\tablenotemark{d} \\
MIPS15880 & 21.26 & -24.17 & 21.74 & 4.22 & 2.83$\times{10^{10}}$ & distorted\tablenotemark{d} \\
MIPS15928 & 19.65 & -25.57 & 19.31 & 2.88 & 1.04$\times{10^{11}}$ & distorted\tablenotemark{c}\\
MIPS15949 & 20.73 & -25.42 & 20.00 & 2.36 & 9.00$\times{10^{10}}$ & none\\
MIPS15958 & 21.09 & -24.83 & 19.25 & 1.43 & 5.19$\times{10^{10}}$ & none \\
MIPS15977 & 19.83 & -25.92 & 19.05 & 2.35 & 1.42$\times{10^{11}}$ & distorted \\
MIPS16059 & 20.49 & -25.98 & 20.19 & 2.82 & 1.51$\times{10^{11}}$ & binary \\
MIPS16080 & 20.31 & -25.70 & 19.62 & 2.43 & 1.16$\times{10^{11}}$ & distorted\\
MIPS16095 & 19.96 & -25.73 & 18.93 & 2.09 & 1.20$\times{10^{11}}$ & none \\
MIPS16113 & 21.40 & -24.42 & 20.94 & 2.71 & 3.58$\times{10^{10}}$ & distorted\tablenotemark{e} \\
MIPS16122 & 21.36 & -24.56 & 21.14 & 3.03 & 4.08$\times{10^{10}}$ & distorted \\
MIPS16144 & 20.85 & -25.27 & 20.80 & 3.24 & 7.83$\times{10^{10}}$ & distorted\tablenotemark{e} \\
MIPS22204 & 20.17 & -25.89 & 18.50 & 1.53 & 1.39$\times{10^{11}}$ & none \\
MIPS22277 & 20.08 & -25.55 & 19.10 & 2.15 & 1.01$\times{10^{11}}$ & none\tablenotemark{b} \\
MIPS22303 & 22.76 & -23.61 & 22.13 & 2.44 & 1.70$\times{10^{10}}$ & none \\
MIPS22530 & 21.07 & -24.83 & 21.88 & 4.87 & 5.25$\times{10^{10}}$ & binary \\
MIPS22558 & 21.62 & -25.58 & 21.39 & 2.71 & 1.03$\times{10^{11}}$ & distorted \\
MIPS22651 & 20.11 & -25.46 & 19.32 & 2.35 & 9.35$\times{10^{10}}$ & none\tablenotemark{d} \\
MIPS22661 & 20.22 & -25.38 & 19.48 & 2.40 & 8.65$\times{10^{10}}$ & none \\
MIPS22699 & 22.19 & -24.45 & 20.88 & 1.75 & 3.67$\times{10^{10}}$ & none \\

\enddata

\tablenotetext{a}
{The $H$-band (or restframe optical) apparent and absolute magnitudes, 
half-light radii, and surface brightnesses were derived using Sextractor, 
which models the galaxies as series of isophotal ellipses. For the spatially 
overlapping binary systems, these numbers are computed for the system as a 
whole. No spectroscopic $k$-corrections have been applied to these quantities.}

\tablenotetext{b}
{In these systems the nucleus does not seem distorted, but secondary low
surface-brightness sources exist within a projected separation of 10 kpc.
Although these systems may be dynamically related, no prominent 
indication of interaction between them can be seen at our surface-brightness 
limits.}

\tablenotetext{c}
{In these cases, the bright nucleus has a faint component in the same envelope
that could either be a secondary nucleus or a bright star-formation knot. }

\tablenotetext{d}
{Bright sources of $\Delta m <$1.2 exist close to these sources but there
are no signs of interaction between them.}

\tablenotetext{e}
{Secondary, low surface-brightness structures exist around the distorted
nucleus of these systems.}


\end{deluxetable}

\begin{deluxetable}{ccccccccccc}
\tablecolumns{11}
\tabletypesize{\tiny}
\tablewidth{0pt}
\tablecaption{\label{tab:decomposition} Bulge-to-disk Decomposition Parameters}
\tablehead{
\colhead{Galaxy} 
& \colhead{$m_H$(bulge) \tablenotemark{a}} 
& \colhead{$\reff$ (bulge) \tablenotemark{a}} 
& \colhead{$m_H$(disk) \tablenotemark{a}} 
& \colhead{$\reff$ (disk) \tablenotemark{a}} 
& \colhead{bulge-to-disk} 
& \colhead{residuals\tablenotemark{b}} 
& \colhead{$n$  \tablenotemark{c}}
& \colhead{$m_H$($n$) \tablenotemark{c}}
& \colhead{$\reff$($n$)}
& \colhead{$m_H$(PSF) \tablenotemark{c}}\\

 & \colhead{} & \colhead{(kpc)} & \colhead{} & \colhead{(kpc)} & \colhead{ratio}
& \colhead{\%}
& \colhead{} & \colhead{} & \colhead{(kpc)} & \colhead{}
}

\startdata
MIPS227   & 19.90 & 2.37 & 19.30 & 1.98 & 0.57 & 6.25  & 1.23 & 18.83 & 2.35 & 21.97 \\
MIPS289   & 20.89 & 1.25 & 20.14 & 5.76 & 0.50 & 3.51  & 0.60 & 20.07 & 3.66 & 21.91 \\
MIPS8196  & 19.15 & 2.06 & 19.64 & 4.36 & 1.57 & 0.97  & 2.65 & 18.64 & 3.10 & 22.14 \\
MIPS15928 & 22.60 & 1.20 & 19.78 & 2.64 & 0.07 & 14.2  & 0.67 & 19.85 & 2.57 & 21.49 \\
MIPS15977 & 20.86 & 1.13 & 20.21 & 2.01 & 0.55 & 1.00  & 0.54 & 20.09 & 2.08 & 21.31 \\
MIPS16095 & 21.15 & 1.38 & 20.36 & 2.03 & 0.48 & 0.72  & 1.35 & 20.41 & 1.94 & 23.31 \\


\enddata
\tablenotetext{a}
{The best-fit solutions for the bulge and the disk magnitudes and half-light radii 
were derived using {\it galfit}. For an exponential disk, the half-light radius is 
related to the scale length $R_d$ as \reff = $1.68 R_d$.}
\tablenotetext{b}
{The residuals are given as a percentage fraction of the initial flux.}
\tablenotetext{c}
{For comparison with the bulge-to-disk decomposition results, we also perform a
free S\'ersic index $n$ versus PSF decomposition. We provide the value of $n$ and 
the magnitude of the S\'ersic component, as well as the PSF magnitude.}
\end{deluxetable}

\begin{deluxetable}{ccccccc}
\tablecolumns{7}
\tabletypesize{\tiny}
\tablewidth{0pt}
\tablecaption{\label{tab:simulations} Photometric parameters of simulated z=2 galaxies}
\tablehead{
\colhead{Galaxy} 
& \colhead{$m_H$(init)  \tablenotemark{a}}
& \colhead{$<\mu_H {\rm (init)}>$}
& \colhead{$\reff$ (init) \tablenotemark{a}}
& \colhead{$m_H$(obs)  \tablenotemark{b}}
& \colhead{$<\mu_H {\rm (obs)}>$}
& \colhead{$\reff$ (obs)  \tablenotemark{b}}  
\\

\colhead{template} & \colhead{} & \colhead{(mag arcsec$^{-2}$)} & \colhead{(kpc)} & \colhead{} & \colhead{(mag arcsec$^{-2}$)} & \colhead{(kpc)} 
}

\startdata

IC~4687         & 19.00 & 20.61 & 7.01 & 19.37 & 20.65 & 6.03 \\
IC~4687         & 20.00 & 21.61 & 7.01 & 20.51 & 21.49 & 5.24 \\
IC~4687         & 21.00 & 22.62 & 7.01 & 21.84 & 21.90 & 3.43 \\
IC~4687\tablenotemark{c} & 22.00 & 23.62 & 7.01 & \nodata & \nodata & \nodata \\

IRAS~09111-1007 & 19.00 & 20.37 & 6.29 & 19.31 & 20.45 & 5.65 \\
IRAS~09111-1007 & 20.00 & 21.37 & 6.29 & 20.53 & 21.30 & 4.77 \\
IRAS~09111-1007 & 21.00 & 22.37 & 6.29 & 21.74 & 22.13 & 4.00 \\
IRAS~09111-1007\tablenotemark{c} & 22.00 & 23.37 & 6.29 & \nodata & \nodata & \nodata \\

IRAS~23128-5919 & 19.00 & 19.70 & 4.62 & 19.20 & 19.63 & 4.06 \\
IRAS~23128-5919 & 20.00 & 20.70 & 4.62 & 20.44 & 20.36 & 3.22 \\
IRAS~23128-5919 & 21.00 & 21.70 & 4.62 & 21.60 & 21.30 & 2.91 \\ 
IRAS~23128-5919 & 22.00 & 22.70 & 4.62 & \nodata & \nodata & \nodata \\ 

NGC~3156        & 19.00 & 17.88 & 1.99 & 19.06 & 17.82 & 1.89 \\
NGC~3156        & 20.00 & 18.88 & 1.99 & 20.10 & 18.78 & 1.82 \\
NGC~3156        & 21.00 & 19.88 & 1.99 & 21.17 & 19.74 & 1.73 \\
NGC~3156        & 22.00 & 20.88 & 1.99 & 22.26 & 20.74 & 1.66 \\

PG~1351+640     & 19.00 & 17.65 & 1.79 & 19.12 & 17.62 & 1.67 \\
PG~1351+640     & 20.00 & 18.65 & 1.79 & 20.17 & 18.56 & 1.59 \\
PG~1351+640     & 21.00 & 19.65 & 1.79 & 21.21 & 19.50 & 1.52 \\
PG~1351+640     & 22.00 & 20.65 & 1.79 & 22.25 & 20.45 & 1.50 \\

\enddata

\tablenotetext{a}
{Magnitude and effective radius of each redshifted source prior to the 
addition of the sky background.}

\tablenotetext{b}
{Magnitude and effective radius of each redshifted source as measured after
the addition of an appropriate sky background. All quantities are measured 
with {\it sextractor}, as for the NICMOS observations.}

\tablenotetext{c}
{These sources cannot be detected at 3$\sigma$ levels.}

\end{deluxetable}

\clearpage

\begin{figure*}
\centering
\includegraphics[width=16cm]{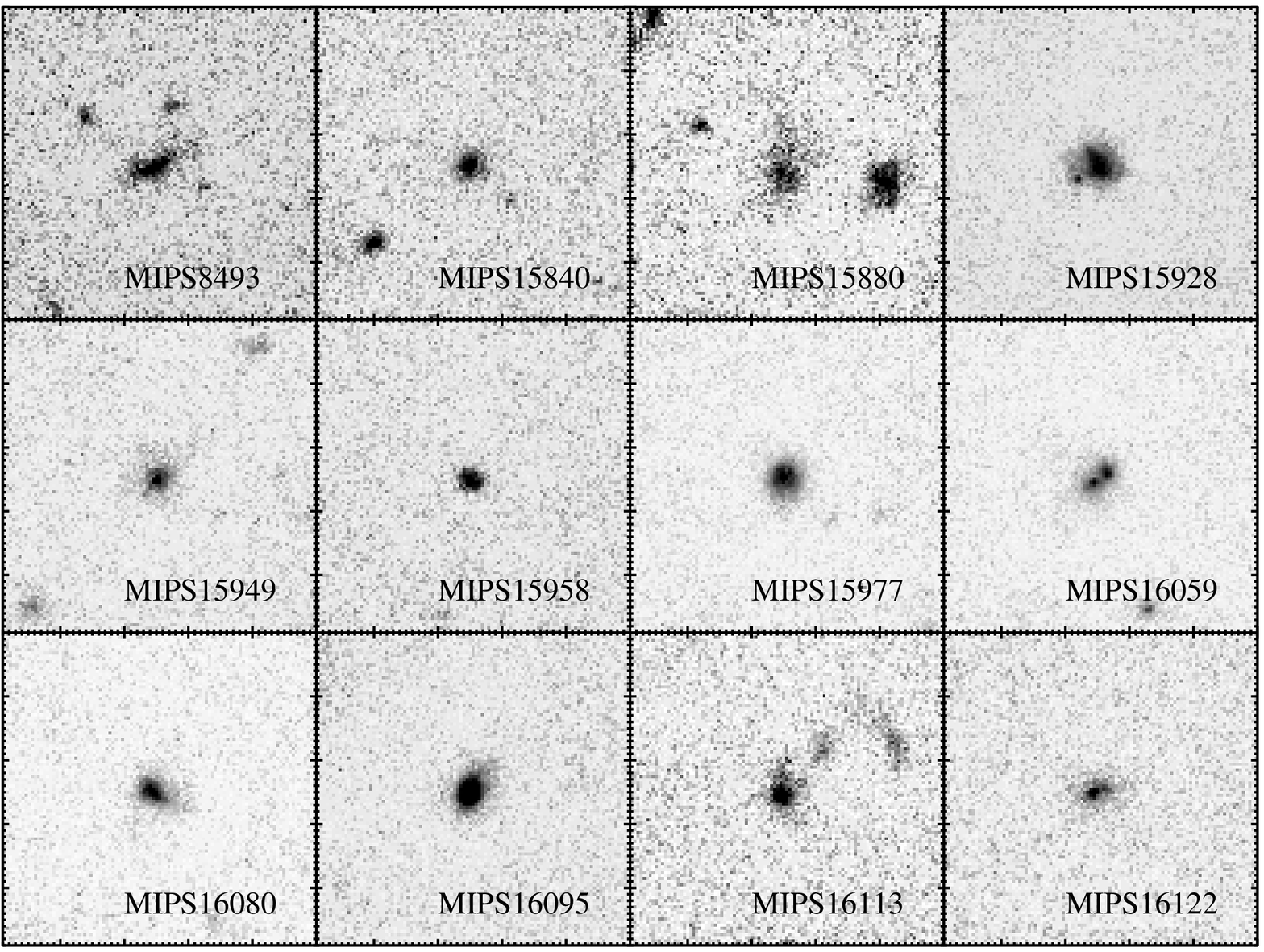}
Fig.~\ref{fig:images}-- continued.
\end{figure*}
\begin{figure*}
\centering
\includegraphics[width=16cm]{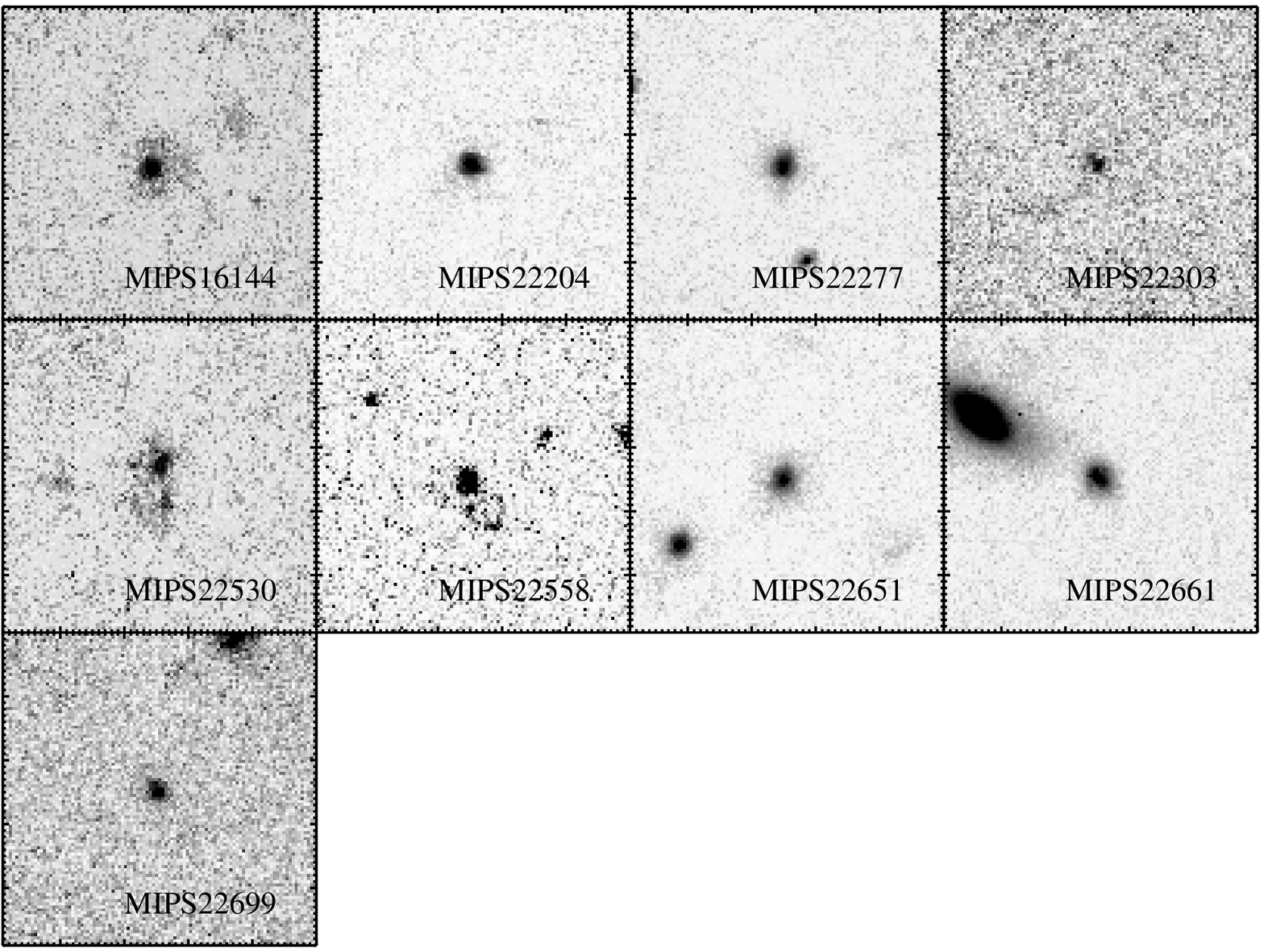}
Fig.~\ref{fig:images}-- continued.
\end{figure*}



\end{document}